\title{Accelerating Lossless Data Compression  with GPUs}

\author{R.L.\ Cloud\thanks{The University of Alabama at Birmingham, rcloud@cis.uab.edu} \and M.L. Curry\thanks{The Universit\
y of Alabama at Birmingham,
curryml@cis.uab.edu} \and H.L. Ward\thanks{Sandia National Laboratories, lee@sandia.gov} \and A. Skjellum \thanks{The Univer\
sity of Alabama at Birmingham, tony@cis.uab.edu}
\and{P. Bangalore \thanks{The University of Alabama at Birmingham, puri@cis.uab.edu}}}

\date{\today}
\documentclass[12pt]{article}

\usepackage[margin=2.54cm, noheadfoot]{geometry} 

\usepackage{graphicx}
\usepackage{url}
\begin{document}
\maketitle

\begin{abstract}
Huffman compression is a statistical, lossless, data compression algorithm that compresses data by assigning variable length codes to symbols, with the more frequently appearing symbols given shorter codes than the less.  This work is a modification of the Huffman algorithm which permits uncompressed data to be decomposed into independently compressible and decompressible blocks, allowing for concurrent compression and decompression on multiple processors.  We create implementations of this modified algorithm on a current NVIDIA GPU using the CUDA API as well as on a current Intel chip and the performance results are compared, showing favorable GPU performance for nearly all tests.  Lastly, we discuss the necessity for high performance data compression in today's supercomputing ecosystem.  
\end{abstract}

\section{Introduction}
Lossless data compression is important in application domains and usage environments where bandwidth or storage limitations may negatively impact application or system performance. Generally classifiable into statistical or dictionary methods, lossless data compression algorithms can range widely in compression speed and efficiency (compression factor). Certain algorithms, especially the more efficient, can be quite computationally expensive, and as the data processing needs of current scientific endeavor continue to scale with more rapidity than storage or bandwidth, compression becomes increasingly necessary, but questions remain as to how to accelerate it and how to do so without consuming the resources devoted to computation.

	The use of graphics processing units (GPUs) for general purpose computation, i.e. problems outside the graphical domain, is a relatively recent development. First this was achieved though third party toolkits, e.g. Stanford's BrookGPU, but even more recently have GPU manufacturers themselves begun to offer general purpose tools which give the programmer a lower level communion with the chip than earlier GPGPU programming interfaces which are built upon OpenGL and DirectX.  One of these, and currently the most prominent, is the Compute Unified Device Architecture(CUDA) from the NVIDIA corporation. The potential benefits of GPUs in general purpose computation are great, but potential must be emboldened, more so even than for parallel programming on the x86. To achieve anywhere near the theoretical maximums in performance  on the GPU, the computation patterns underlying a solution's algorithm must be very near to the traditional usage of the GPU.  A prospective algorithm's implementation on the GPU should be, in order of importance to performance, highly data parallelizable, logically simple, and have relatively many computations to memory accesses. In essence, to use the GPU to maximum effect, the abstractable computation patterns underlying a solution should be co-linear to the GPUs original task, graphics rendering. Our problem domain, I/O, while it does not perfectly fit these criteria, has already benefited from GPUs to enhance storage redundancy~\cite{mlcurry}; we attempt now their utilization in lossless data compression.
	
	One major difficulty here in achieving good speedup with slim negative side effects is that lossless data compression algorithms can generally not be, in their unaltered form, thought of as highly parallelizable. Indeed, if one wishes to express these algorithms in parallel, one often needs to consider tradeoffs between compression efficiency and performance. Nevertheless, we hope to effectively demonstrate that it is possible to come to a reasonable middle ground with respect to coding acceleration and efficiency loss.

\section{Huffman Compression}\label{huffman}
Statistical methods of data compression perform analysis on the nature of the data to make intelligent decisions about how it can be represented more efficiently in compressed form. The Huffman encoding algorithm falls within this genus and operates by counting the appearance of every distinct symbol in the uncompressed data, then representing the more frequent with shorter codes than the less. Every symbol in the data is replaced with its code, and if the data is non-random, i.e.  a few symbols appear with greater frequency than others, compression can be achieved. The Huffman compression algorithm is old by the standards of our science~\cite{huffman}, but is still used, and has the attractive quality of being a primitive of several more modern and common algorithms, e.g. Deflate~\cite{gzip} and potentially the algorithm described by Burrows and Wheeler~\cite{bwt}.

\subsection{Parallel Huffman Compression}\label{p_huffman}
There is literature on parallel Huffman coding and of variety, ranging from the actual construction of Huffman codes in parallel~\cite{berman}, ~\cite{atallah} to ~\cite{klein} which addresses details of decomposition for parallel Huffman decoding and demonstrates some moderate decoding speedups while maintaining optimally encoded data by making use of the observation that Huffman codes can frequently synchronize.  Because of limitations in our architecture, we must try to create the simplest encoding routine possible.  In doing this we make a minor modification to the output of the Huffman algorithm.

	An alteration is necessary because of the nature of Huffman codes, i.e. they are of a variable length;  an encoded data string is composed of these codes packed together in a nature where bit codes can cross byte boundaries.  Simple decomposition of the encoded data stream into blocks of static size would result in the practical certainty that decoding would take an erroneous path, which is discussed in some detail in~\cite{klein}. One counter to this is to pack the blocks to byte boundaries, introducing some size overhead. One more change is necessary. Because the codes are of variable lengths, even if we encode a constant number of symbols in each block, the resulting length of the encoded block will vary, sometimes dramatically. For this reason, we must encode an indication of where the block starts and ends. Our approach is again simple; at the start of the encoded block we give the length of the block which is known by making an additional pass over the unencoded block and summing the lengths of the code representation of the symbols.  Our implementation stores this length as an unencoded four byte integer for simplicity, and because of this and the requirements of our architecture, we pack the blocks to four byte boundaries.

\begin{figure}
\centering
\includegraphics[scale=.5]{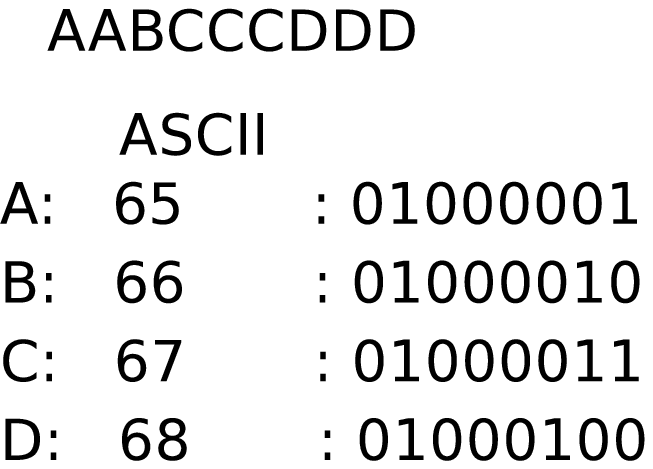}
\caption{The original string and its ASCII representation.  As ASCII encodes each character with a single byte, the nine character string is 72 bits.}
\label{fig:string}
\end{figure}

\begin{figure}
\centering
\includegraphics[scale=.5]{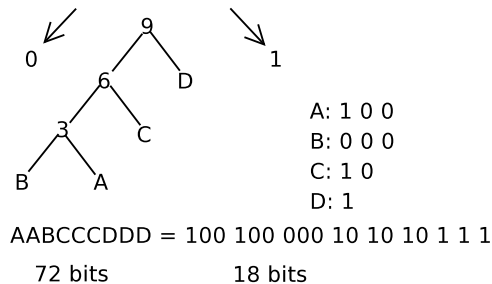}
\caption{The binary tree created by the Huffman algorithm and the encoded representation of the original string.  The encoded representation of a character is 
found by traversing the tree, assigning a binary 0 or 1 for a left or right traversal respectively.}
\label{fig:codes}
\end{figure}

\begin{figure}
\centering
\includegraphics[scale=.5]{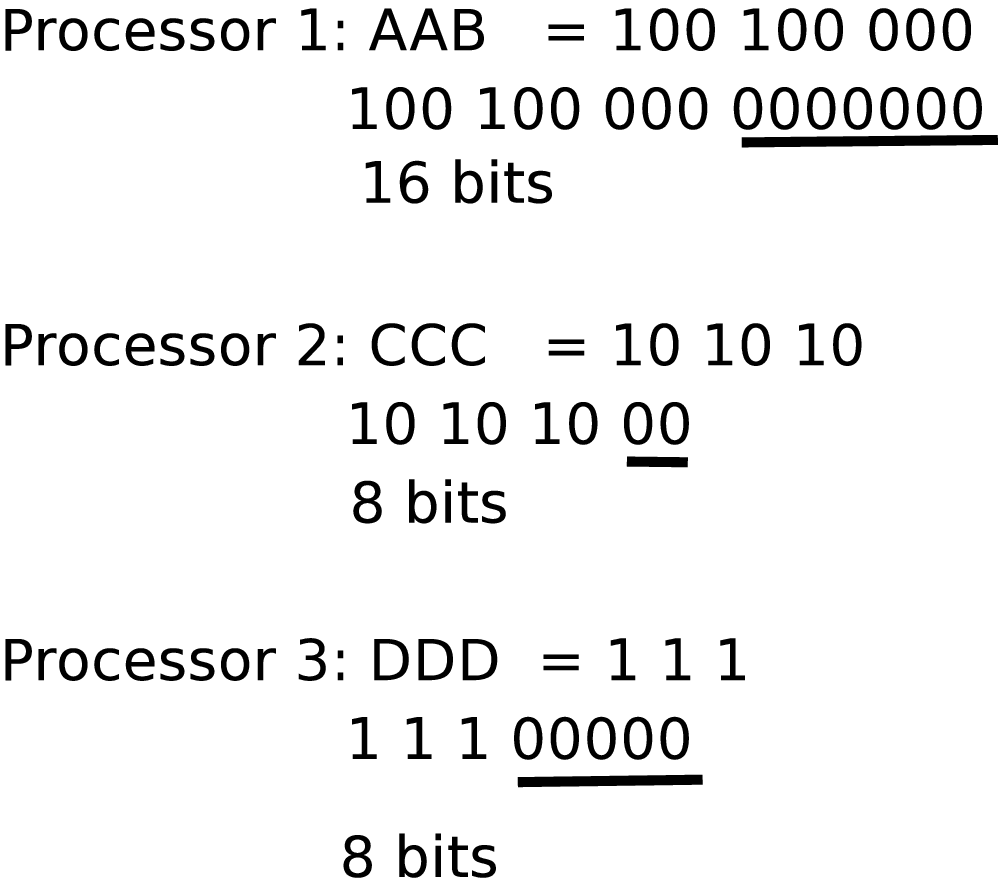}
\caption{Decomposing the string into three symbol blocks and then packing the encoded bits to the next byte boundary.}
\label{fig:packed_bytes}
\end{figure}

\begin{figure}
\centering
\includegraphics[scale=.5]{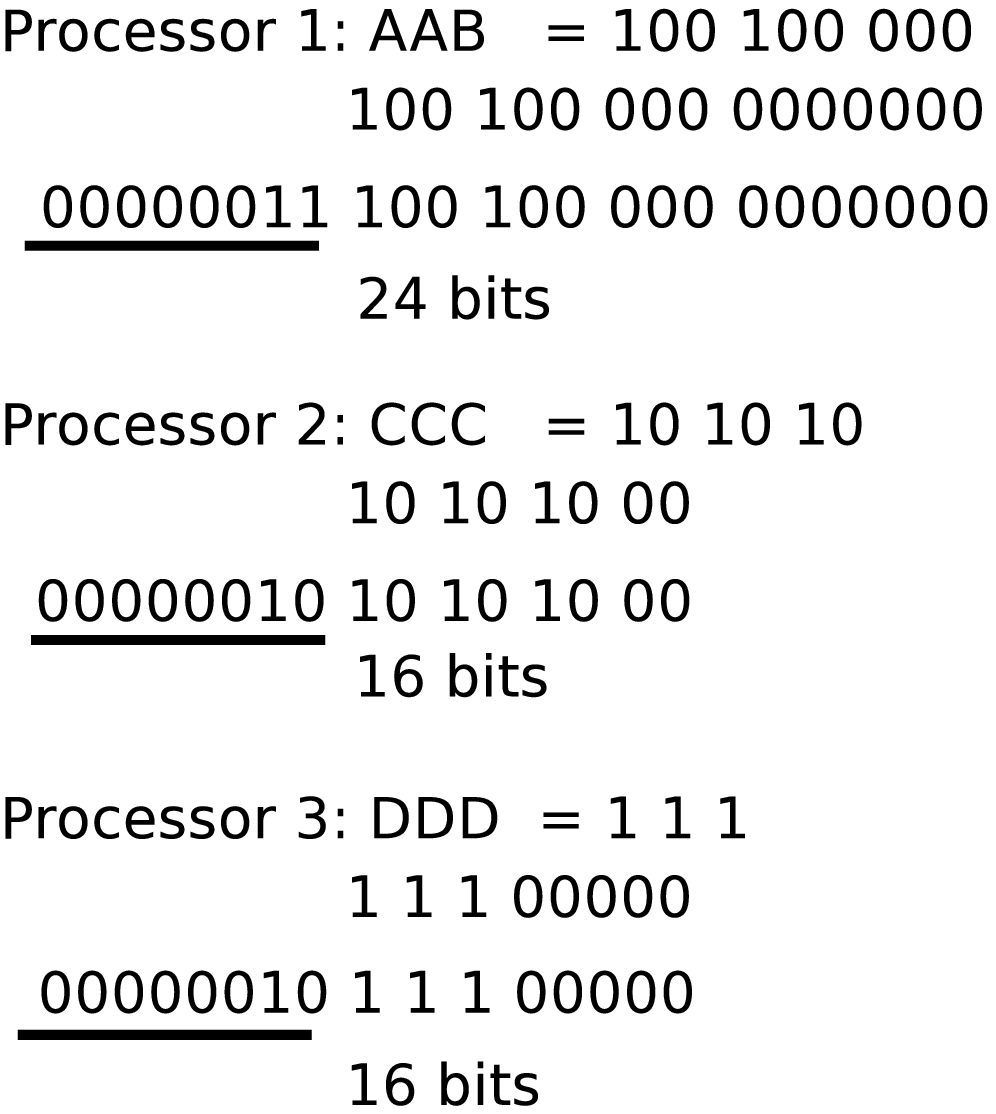}
\caption{The addition of a (underlined)length delimiter at the start of the block.  Single bytes are used for the overhead in the diagrams for simplicity. In our implementation, we pack the block to four bytes and use a four byte integer to represent the block length}
\label{fig:delimiters}
\end{figure}

	The overhead of our modifications range therefore, from between 32 bits and 63 bits per block, with the variation being because if the size of the encoded block is evenly divisible by four bytes, it is unnecessary to add packing bits to its tail. This overhead naturally becomes less significant as the length of the block is increased, which is indicated in the figure measuring block size against overhead. The time required for summing the block lengths is measurable but undramatic and most noticeable when comparing the runtimes of a sequential block encoder to a sequential traditional(non-block) encoder.
	
	To parallelize decoding, it is sufficient to build a table of offsets into the encoded data from these block length delimiters. The computation threads on the GPU can then index into the encoded data and decode in parallel, storing decoded data in a sequential array indexable by thread and block numbers.

\begin{figure}
\centering
\includegraphics[scale=1]{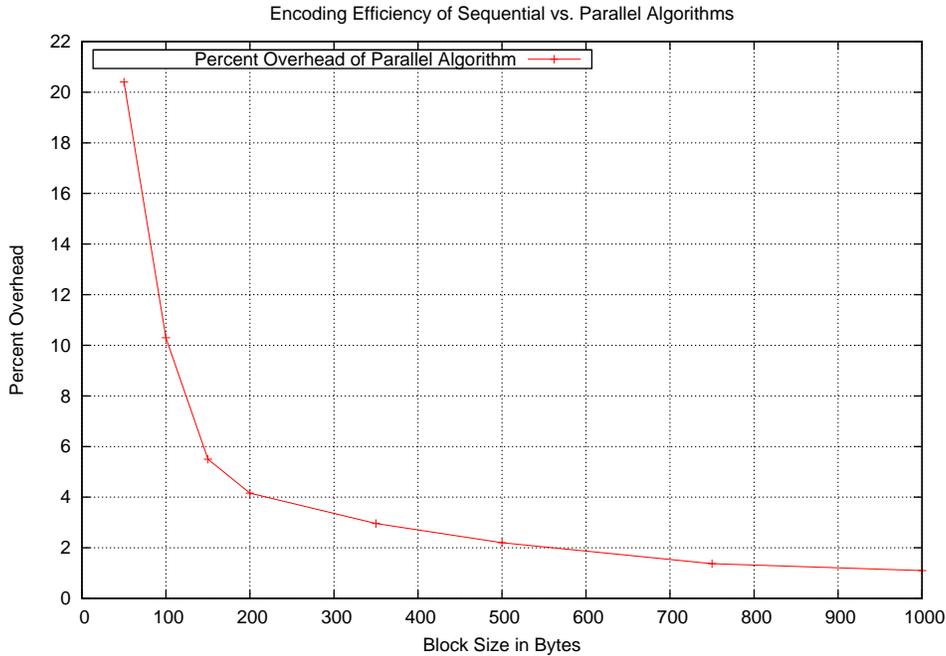}
\caption{The size overhead of using the parallel Huffman algorithm graphed
against the block size. The number of bytes overhead per block remains a constant, so as the block size increases the overhead becomes less significant. At large block sizes, the overhead per block can be less than one percent.}
\label{fig:eff_lost}
\end{figure}

\section{Performance Comparisons}\label{performance_comp}
\subsection{Encoding}\label{perm_comp_enc}
Acceleration over our sequential implementation was achieved for both encoding and decoding.
This comparison is most meaningful in terms of throughputs, the amount of data which can be encoded or decoded per second.  Following is the comparison of our sequential encoder to our parallel GPU encoder and a parallel CPU encoder programmed with OpenMP.   The GPU used in these experiments is the NVIDIA GeForce GTX 285 with 240 cores at 1.5 GHz, and the CPU used is the Intel Core i7 Extreme Edition 965 with four cores at 3.2 GHz.  Despite the GPU having 60 times the number of cores as our CPU, the differences in throughput between the GPU encoder and the OpenMP encoder are not dramatic.  This paradox can be largely resolved by recalling that the architecture of the GPU was developed for the SIMD, single instruction multiple data, programming model while our CPU was developed with MIMD, multiple instruction multiple data, in mind.  

	The processors in the GPU are organized into 30 groups of 8 cores.  Each group of cores is known as a multiprocessor and contains a single control unit and a small amount of high speed memory shared between the cores in the multiprocessor.  The control unit broadcasts an instruction to all the cores, and optimal performance can only be achieved when every core can execute it.  If, for example, the instruction is a branching statement, then there is a likelihood that some cores will not follow the jump, and in this case, some cores must remain inactive until they either themselves satisfy the branching instruction or control passes beyond the branching sections of the code.  Therefore, in the worst case, when only one core can satisfy the jump and the other seven are left idle, our GPU behaves more like a crippled 30 core shared memory MIMD machine with a slow clock speed and no automatic memory caching.  Our encoder consists of complicated branching statements for the bit manipulation which makes worst case behavior relatively likely.  This also illustrates that in heterogeneous programming environments, one must be very aware of the strengths and weaknesses of the various architectures so that programming effort can be directed where benefits are most likely to be found.

\begin{figure}
\centering
\includegraphics[scale=1]{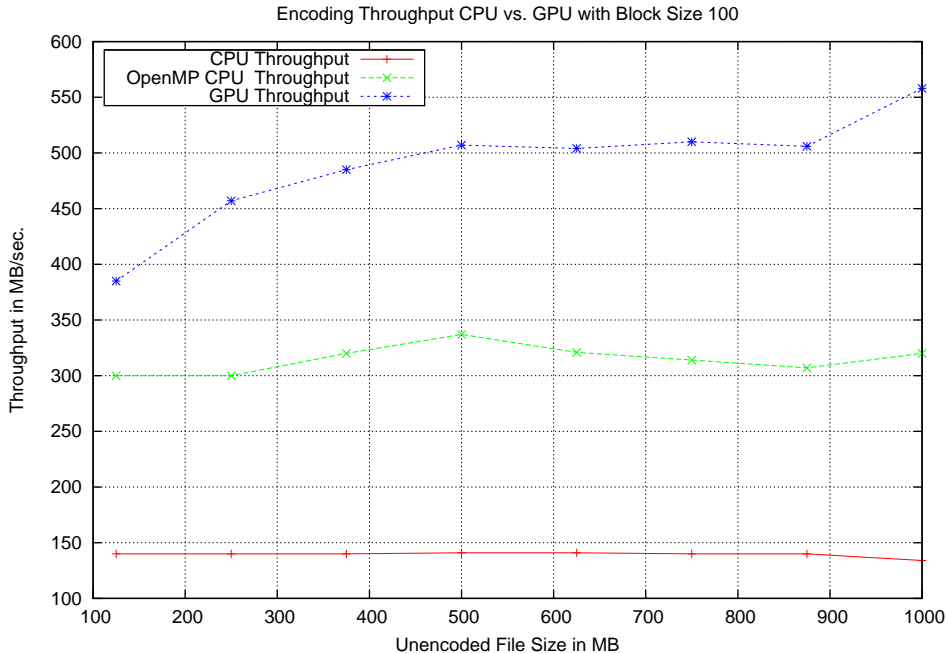}
\caption{We saw superior performance with the GPU based encoder compared to our multi-core CPU encoder and our single threaded CPU implementation}
\label{fig:enc_perf}
\end{figure}

\subsection{Decoding}\label{perm_comp_dec}
Our decoding routine consists of reading bits and traversing a binary tree repeatedly for each code string.  This contains branching instructions, but markedly fewer than the encoding routine, and the factor of acceleration on the GPU is greater than that of the encoding routine.  Also interestingly, the measured increases in throughput from using OpenMP on the CPU, compared to the sequential implementation, are even better than linear by number of cores on the CPU.  By launching increasing numbers of threads, we can hide latency by issuing more memory requests.  In this way, we saw continued performance improvements through increasing thread counts up to 8.  Intel's Hyper-Threading technology assists significantly in this.

\begin{figure}
\centering
\includegraphics[scale=1]{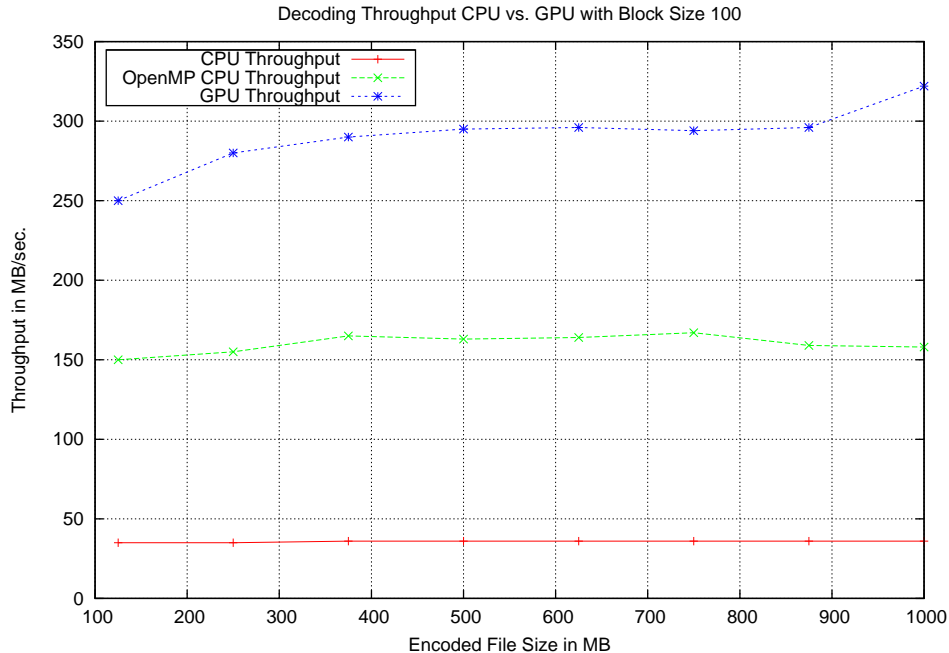}
\caption{Again, our GPU based decoder gave better performance than both CPU decoders.}
\label{fig:dec_perf}
\end{figure}

\section{Conclusions}\label{conclusions}
The data presented here suggests that the strengths of the GPU architecture are robust enough to give performance benefits to applications which, while data parallel, still have a not insignificant level of logical complexity.  Optimal use of the GPU's SIMD cores requires the complete elimination of divergence within warps, which, in practicality, requires the complete absence of if statements from the GPU sub-routine; however, sub-optimal performance, through the emulation of MIMD, can still be acceptable.  Despite the large number of divergent threads in a warp; our encoder kernel is capable of throughputs, sans memory transfer times to and from the GPU, in excess of 4 GB/sec.  Total encoding throughputs using the GPU are weighed down by the need to transfer data to and from the card; however, in an online system, or when encoding very large amounts of data, this could be somewhat ameliorated by using asynchronous data transfers with the GPU to fully exploit bus resources while encoding.

	Realistically, current performance levels for our GPU encoder and decoder do not warrant the use of the program as a standalone encoding system.  The Huffman algorithm itself is not the best choice for such purposes and even the strengths of the GPU do not make up for the algorithm's deficiencies.  However, our encoding system could be used as an auxiliary process to a GPU application.  Much greater coding performance than that shown in the above figures could be seen were the data to be encoded already on the GPU.

%retry experiment again with 32 bit ints. 

%parallel bzip: http://compression.ca/pbzip2/% 
% fermi whitepaper: http://www.nvidia.com/content/PDF/fermi_white_papers/NVIDIAFermiArchitectureWhitepaper.pdf
\bibliographystyle{abbrv}
\bibliography{huff_report}

\end{document}